\documentstyle[12pt,aps,prc,epsfig,preprint]{revtex}
\tightenlines
\begin{document}

\begin{titlepage}
\title{
Elastic scattering of low energy pions by nuclei and the in-medium
isovector $\pi N$ amplitude}
\author{E. Friedman$^{a}$, M. Bauer$^{b}$, J. Breitschopf$^{b}$,
H. Clement$^{b}$, H. Denz$^{b}$, E. Doroshkevich$^{b}$,
A. Erhardt$^{b}$,  G.J. Hofman$^{c}$, S. Kritchman$^{a}$,
R. Meier$^{b}$, G.J.Wagner$^{b}$, G. Yaari$^{a}$\\
$^a${\it Racah Institute of Physics, The Hebrew University, Jerusalem 91904,
Israel\\}
$^b${\it Physikalisches Institut, Universit\"at T\"ubingen, 72076 T\"ubingen,
Germany\\}
$^c${\it TRIUMF, Vancouver, British Columbia, Canada V6T 2A3
and University of Regina, Regina, Saskatchewan, Canada S4S-0A2\\}}

\maketitle
\begin{abstract}

 Measurements of elastic scattering of 21.5 MeV $\pi ^\pm$
by Si, Ca, Ni and Zr were made using a single arm magnetic
spectrometer. Absolute calibration was made by parallel measurements of 
Coulomb scattering of muons. Parameters of a pion-nucleus optical potential
were obtained from fits to all eight angular distributions put together.
The `anomalous' $s$-wave repulsion known from pionic atoms is clearly 
observed and could be removed by introducing
a chiral-motivated density dependence of the isovector scattering amplitude,
which also greatly improved the fits to the data.
The empirical energy dependence
of the isoscalar amplitude also improves the fits to the data
but, contrary to what is found with pionic atoms, on its own is incapable 
of removing the anomaly.

\end{abstract}



Corresponding author: E. Friedman, 
Tel: +972 2 658 4667, FAX: +972 2 658 6347, \newline \indent
E mail: elifried@vms.huji.ac.il

\centerline{\today}
\end{titlepage}

\section{Introduction}
\label{sec:int}

Interest  in the pion-nucleus interaction 
at low energies
has been focused recently on the $s$-wave part of the pion-nucleus 
optical potential.
The so-called `anomalous' repulsion of the $s$-wave pionic atom potential
is the empirical finding, from fits of optical potential
parameters to pionic atom
data, that the strength of the repulsive $s$-wave
potential inside nuclei is nearly double the value expected on
the basis of the free $\pi N$ interaction. 
This  has been known for almost two decades \cite{BFG97}.
The enhancement results mostly
from the  in-medium
isovector $s$-wave $\pi N$ amplitude
which plays a dominant role due to the nearly vanishing of the corresponding
isoscalar amplitude. Some extra repulsion comes also from the empirical
dispersive component of the two-nucleon absorption term.
The renewed interest in pionic atoms is based partly on the 
experimental observation  of `deeply
bound' pionic atom  states in the (d,$^3$He) reaction
\cite{YHI96,GGK00,GGG02,SFG04}, the existence of which was
predicted a decade earlier
\cite{FSo85,TYa88,THY89}. It is also based partly 
on attempts to explain the anomalous
$s$-wave repulsion
 in terms of  density dependence of the pion decay constant
\cite{Wei01}, or  by
constructing the $\pi N$ amplitude near threshold within a systematic
chiral perturbation expansion \cite{KWe01,MOW02}
and in particular imposing on it gauge invariance \cite{KKW03,KKW03a}.

Large scale fits to pionic atom data encompassing the whole of the 
periodic table showed \cite{Fri02,Fri02a} that indeed the 
density-dependence of the pion decay constant \cite{Wei01} which
causes the {\it isosvector} scattering amplitude to become density dependent,
is capable of removing the anomaly. Similar conclusions were 
also presented
\cite{KYa01,GGG02,GGG02a,SFG04} on the basis of very restricted data 
sets, consisting
mainly of the recently observed `deeply bound' states of pionic atoms.
Such analyses of small data bases 
inevitably rely on assumptions regarding some of the parameters
of the potential, a method which also leads to unrealistic estimate 
of errors, as 
demonstrated \cite{FGa03} in a comparative study of uncertainties. It is
interesting to note in this context that the deeply bound states have 
failed, so far, to provide information which is not available from the host
of data regarding normal states. This is fully understood \cite{FGa98,FGa03a}
from arguments of overlap between the pionic wavefunction and the nucleus.

In an alternative approach large scale fits
showed \cite{FGa04} that the anomaly can be removed
by imposing the minimal substitution requirement \cite{ETa82} of
$E \to E - V_{c}$, where $V_{c}$ is the Coulomb potential,
on the properly constructed pion optical potential.
In this case the removal of the anomaly  is essentially thanks
to the empirical energy dependence of the {\it isoscalar} scattering
amplitude. 
This state of affairs makes it highly desirable to extend the experimental
basis for studies of the pion-nucleus interaction at low energies and this 
is the topic of the present paper.

In the present work we  extend the study of the $s$-wave
term of the pion-nucleus potential by considering the elastic scattering
of very low energy $\pi^+$ and $\pi^-$ on several nuclei.
With the large number of scattering experiments performed in the first
two decades of the `pion factories', it
is somewhat surprising to realize that at
kinetic energies well below 50 MeV there seems to be only one set of
high quality
data available for both charge states of the pion obtained in
the same experiment,
namely, the data of Wright et al. \cite{Wri88} for 19.5 MeV pions on calcium.
We have therefore performed precision measurements of elastic scattering
of 21.5 MeV $\pi^+$ and $\pi^-$ on several nuclei in order to provide
the necessary data.
The purpose of this experiment is
to study the behavior of the pion-nucleus potential across
threshold into the scattering regime  and  to examine
if the above-mentioned anomaly is observed also above threshold.
Of particular importance is
the question of whether the density dependence 
of the isovector amplitude or the empirical energy
dependence of the isoscalar amplitude, 
which remove the anomaly in pionic atoms, are required 
by the scattering data. In the scattering scenario, unlike in the atomic
case, one can study both charge states of the pion, thus increasing
sensitivities to isovector effects and to the energy dependence 
of the isoscalar amplitude due to the Coulomb interaction.
The present paper reports on precision measurements of elastic scattering
of 21.5 MeV $\pi ^+$ and $\pi ^-$ by several nuclei and on the analysis of
the data within an optical potential approach. A first account
of this work has already been published \cite{FBB04}.

Section \ref{sec:exp} describes the experimental set-up and section
\ref{sec:data} describes the data reduction and summarizes the
experimental results. Section  \ref{sec:analysis} presents the 
analysis in terms of several versions of the pion-nucleus optical
model. Section \ref{sec:sum} is a summary.

\section{Experiment}
\label{sec:exp}
\subsection{Experimental set-up}
The aim of the experiment was to measure the differential cross
sections for elastic $\pi^+$ and
$\pi^-$ scattering off Si, Ca, Ni and Zr with  sufficient energy
resolution to separate the first excited states of the target nuclei
and with high absolute accuracy. This was achieved with a single-arm
experiment using a magnetic spectrometer which allowed scattering
angles up to $120^{\circ}$ in the lab system.

The experiment was performed at the low-energy pion channel $\pi$E3 
\cite{For97} of the Paul-Scherrer-Institute (PSI) at Villigen,
Switzerland in its chromatic mode. The channel was designed to match
the optical 
characteristics of the low-energy pion spectrometer LEPS that we used
(see below). It is an S-type channel with a vertical bending plane
resulting in an experimental area 6 m above floor level.
It has a momentum acceptance of 3.4\% FWHM , a momentum resolution
of $<0.3\%$ and a length of 13 m. It focussed the pions on a beam
spot of about 5 by 10 cm$^2$  on the target with a vertical dispersion
of $-$5 cm/\%. The target was mounted in air about 1 m
downstream of the last quadrupole magnet of the channel and separated
by thin Mylar windows from the vacua of the channel and 
of the spectrometer.

The dispersed beam spot demanded large area targets. 
We used self-supporting targets of Si, Ca, Ni and Zr of natural
isotopic composition which were hanging on thin threads over the pivot
point of the spectrometer. The target heights and widths ranged from 
10 by 15 cm$^2$  for Si to 27 by 32 cm$^2$ for Zr and their
areal densities amounted to 213, 325, 178 and 160 mg/cm$^2$ for Si,
Ca, Ni and Zr, respectively.
While Ni and Zr were simply chemically pure metallic plates the Si
target was glued together from two layers of 5 by 5 cm$^2$ wafers. The
Ca target consisted of 12 pieces of  $\approx$ 4.4 by 4.8 by 0.2 cm$^3$, 
cut from a block of Ca and also glued together with tiny amounts of glue.
To minimize the energy loss and straggling of the pions in the targets their
angle settings $\theta_{tgt}$ were always chosen such that the normal
formed an angle of $\theta_{tgt}=\theta_{lab}/2$  with the beam direction
where $\theta_{lab}$ is the spectrometer angle in the laboratory system.

The low-energy pion spectrometer LEPS \cite{MGJ87,JMJ95} consists of 
two dipole magnets
in a split-pole configuration and a symmetric triplet of quadrupole
magnets placed in front of the dipoles\footnote{The spectrometer was
dismantled after this experiment}. In the intermediate focus 
a multiwire proportional chamber consisting of three vertical and
three horizontal readout planes allows the measurement of position and
angles of the trajectory. The focal plane is tilted by 43$^{\circ}$ with
respect to the central trajectory. This permits the use of a vertical
drift chamber to determine the coordinate along the dispersive
direction and the corresponding angle in the focal plane.
The drift chamber is followed by two 
scintillation counters, each 10 mm thick, used for trigger purposes 
and by a range telescope consisting of six scintillation detectors of 15
mm thickness each, which is used
for particle identification. More details are given in
\cite{JMJ95}. However, in addition to the  set-up used
previously we employed a
thin (2 mm) plastic scintillator detector in front of the quadrupole
triplet that supplied a precise timing signal which moreover is
free of ambiguities, in contrast to the
cyclotron RF-signal used in \cite{JMJ95}. 
The resulting loss in energy resolution was
insignificant relative to that produced by straggling in the targets,
and the precise time-of-flight (TOF) measurement through the spectrometer
improved the particle identification (see below).
 
The absolute normalisation of the cross sections for pion scattering
is based on the so-called {\it lepton normalisation} developed for use
with the LEPS-spectrometer  \cite{BKC90}. In essence, the cross
sections are obtained by measuring the yields 
of pions relative to elastic scattering of muons
where the corresponding cross sections may be reliably
calculated from the known charge distributions of the target nuclei.
This way, quantities such as target thicknesses and effective solid
angles play no role. The method employs a monitoring system \cite{BKC90}
consisting of a muon telescope ring upstream of the target 
and a hodoscope placed 40 cm downstream of the target. 
The hodoscope consists of two planes of scintillator strips, 2 mm thick,
in x- and y-direction, respectively, which allow 
TOF and intensity
measurements over an area of 24  by 12 cm$^2$ with a pixel size of 
1.5  by 1.5 cm$^2$. A 2 mm sheet of aluminum in front shielded the
hodoscope against low-energy protons from the channel. The muon ring
has been devised as a secondary low-rate beam monitor measuring the decay
muons from pion decay in flight. It consisted of four scintillator
telescopes mounted in a ring ({\it up, down, right, left})
around the incident beam direction with a
directional sensitivity centered about 17$^{\circ}$, well within
 the muon decay cone. Each of these passing
telescopes consists of a pair of scintillators 70 by 15 by 
3 mm$^3$ coupled to miniature phototubes.

Finally, typical pion beam intensities were 0.5$\times 10^6$ sec$^{-1}$
for $\pi ^+$ and 0.1$\times 10^6$ sec$^{-1}$ for $\pi ^-$.
The data acquisition system essentially remains as described in\cite{JMJ95}.

\subsection{Experimental procedure}

\begin{figure}
\epsfig{file=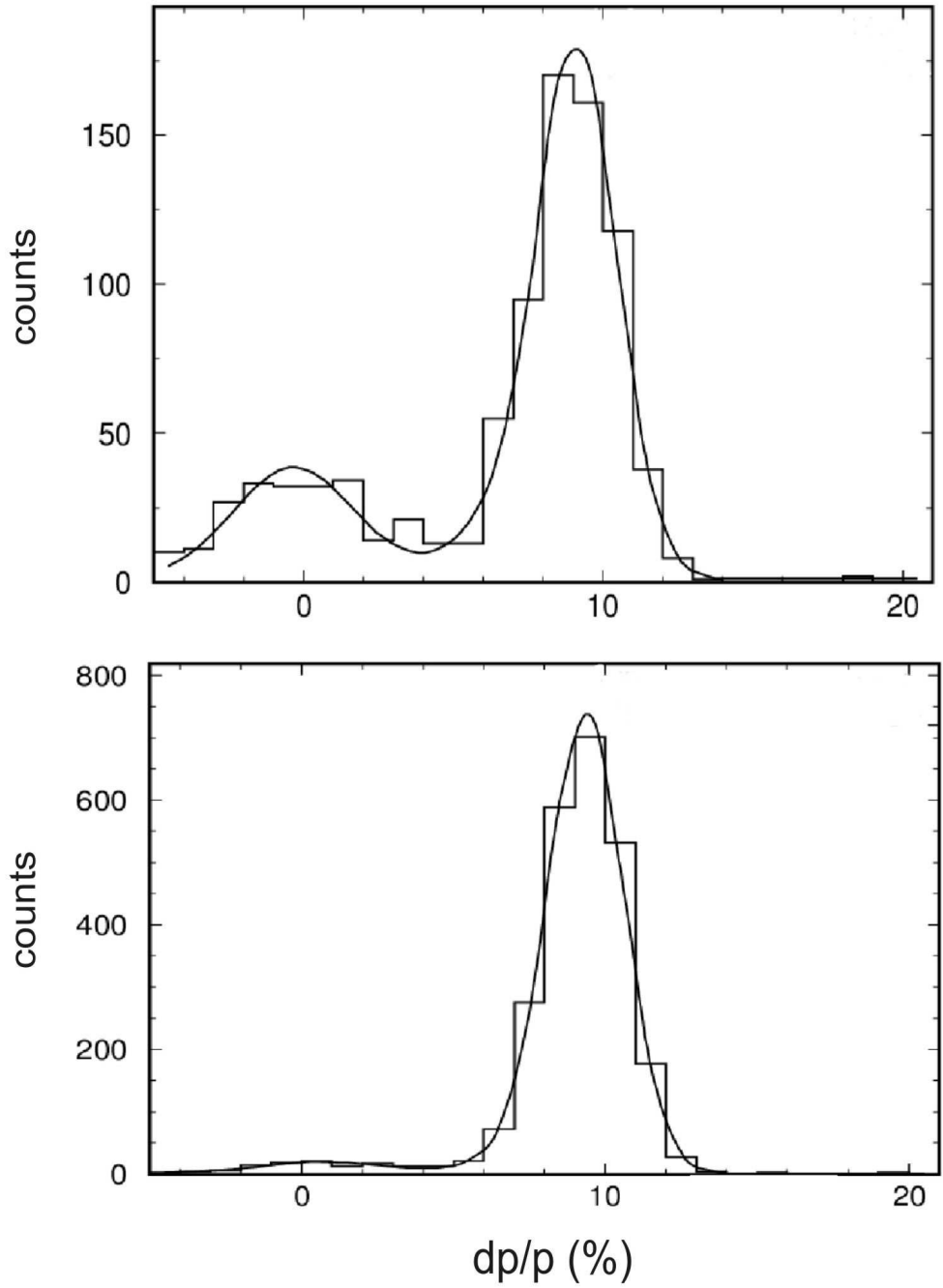, height=190mm,width=150mm}
\caption{Examples of focal-plane spectra for $\pi ^-$  scattered
by Zr at 110$^\circ$ (upper part) and at 80$^\circ$ (lower part). 
A peak due to 
inelastic scattering is clearly seen at 110$^\circ$.}
\label{fig:fpspec}
\end{figure}

The $\pi$E3 channel was set to a momentum of 82.4$\pm$0.3 MeV/c based on the
momentum calibration of \cite{JMJ95}. The magnetic fields of the 
LEPS-spectrometer were set to
a nominal value of 69.4 MeV/c and the accepted momenta
in its focal plane were in the range
$-$15$\%\le {\Delta p /p} \le+20$\%, where $\Delta p$ denotes
the deviation from the central momentum. The energy resolution was about 
0.4 MeV ($1 \sigma$), enough to separate off the first excited states of
all target isotopes with the exception of $^{43}$Ca and $^{61}$Ni
which, however, have an isotopic abundance of only 0.14\% and 1.14\%,
respectively. 
Figure \ref{fig:fpspec}  shows examples of focal plane
spectra for $\pi ^-$ scattered from the Zr target at 80$^\circ$ and
110$^\circ$. In the latter case inelastically scattered pions
are also observed. For the other targets and for most angles there was
hardly any inelastic scattering observed.

\begin{figure}
\epsfig{file=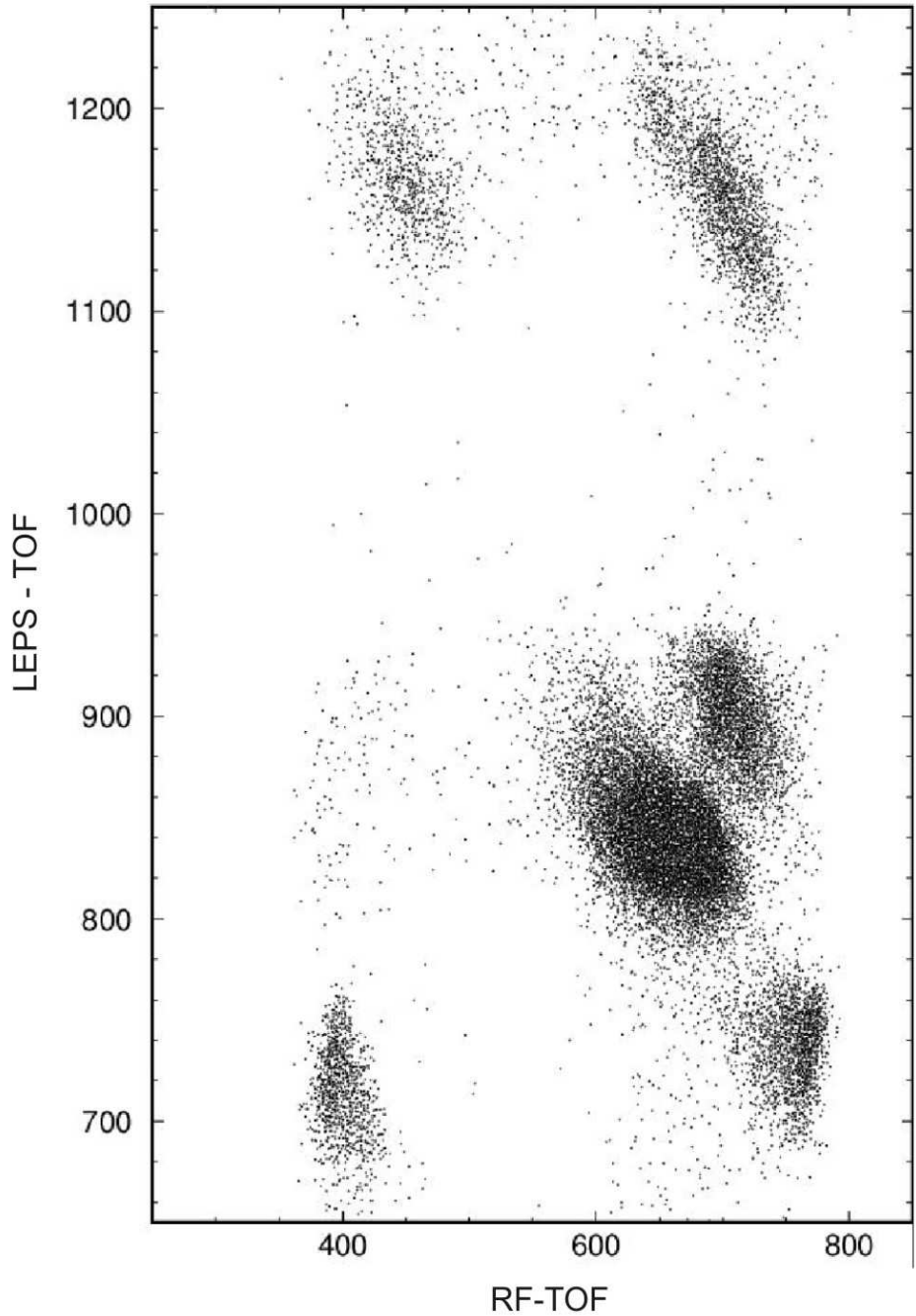, height=190mm,width=150mm}
\caption{TOF within the LEPS spectrometer {\it vs.}
TOF relative to the cyclotron RF, in arbitrary units, 
showing clear separation of the 
various types of particles. Pions are at the bottom and electrons are at the
top of the figure. The central groups are due to beam muons (right) and
decay muons.}
\label{fig:tof}
\end{figure}

Unambiguous particle identification  in the focal plane was
made possible in a 2D representation
 of TOF through the spectrometer
{\it vs.}
the TOF relative to the cyclotron RF signal,
see Fig. \ref{fig:tof}. The splitting of the RF-TOF for the pions
at the bottom of the figure (and for electrons at the top)  had no
consequences for the measurements as both parts of the pion peak 
were always summed.
In  addition to the {\it loci} of scattered pions, muons and electrons the
signals from muons resulting from pion decay in the spectrometer were
clearly discernible. The 2D analysis is essential as 
about two thirds
of the pions leaving the target decayed before reaching the focal
plane, producing a sizeable muon background there. 

For the {\it lepton normalisation} it is essential to know the
{\it relative} acceptance of the spectrometer which varies along the
focal plane by over a factor of two  \cite{BKC90}. The relative acceptance 
was measured by
performing a ``momentum scan'' where the normalized yield of elastic 
scattering was  measured as a function of
the momentum setting of LEPS, for a fixed scattering angle and
beamline tune. The relative normalisation was made both with the 
 muon telescope ring and  the beam hodoscope described above.
In order to use the muon normalisation technique it is necessary to
know the ratio between muon and pion beam intensities at the
scattering target. The ratio on the hodoscope is available
for each pixel and it could be extrapolated back to the target
position with the help of Monte Carlo calculations, taking into account that
muons which originate from pion decays near the hodoscope are still
being identified by TOF as pions. The calculated ratios 
varied by less than $\pm$0.7\% when varying the initial conditions
of the calculation or  the target material or angle. As a result of
these simulations
the $\mu / \pi$ ratios measured on the hodoscope had to be divided
by 1.040$\pm$0.015 in order to get the ratios of genuine muons to pions
on the target.

Two types of measurements of elastic scattering of muons were made.
 In the first type (`self'
measurements), muons were
recorded simultaneously with the pions, which was possible because
muons appeared in the focal plane with slightly higher momentum 
than pions due to the different energy losses of pions and muons in
the target, in the scintillation counter in front of the quadrupole
triplet and in the vacuum windows. The relative yields were then corrected for
the acceptances at the two locations.  In addition and immediately
after some pion measurements, designated `muon runs' were made 
 where the magnetic fields of LEPS were scaled up by 4.7\% which 
moved the muons in the focal plane to the previous location of the
pions. In this method there was no need to take into account the position
dependence of the acceptance. However, this method was not practical at
the largest angles due to the small cross sections. 
As is reported below, both methods yielded the
same average normalisation constants within 2\%, whereas the estimated errors 
of the `self' and of the `muon runs' methods were 7\% and 5\%, respectively.

\section{Data reduction and experimental results}
\label{sec:data}
Data reduction can be described as a series of cuts performed to finally
define the events which are used to derive the required cross sections.
The first cut was made in the focal plane angle {\it vs.} the intermediate
focus angle, which removed many of the muons that originate from pion
decays within the spectrometer. After such a decay the simple geometrical
correlation between the two angles is lost and 
consequently this cut removed many of the 
decay muons, which at such low energies originate from close to 70\% of the
pions entering the spectrometer. The second cut was made in the 2D TOF
spectra, such as shown in Fig. \ref{fig:tof}. The boundaries of the peaks 
were defined manually for each target and for every angle.
In this cut elastically scattered pions or muons were selected, the latter
related to beam muons (in contrast to decay muons) which served for the
normalisation, as described above. The third cut was made on the coordinates
at the target, defining a fixed area on the target from which particles 
were collected by the spectrometer. 
These coordinates were calculated by reconstructing the trajectory for each
particle using the chamber readings at the focal plane and at
the intermediate focus. The calculated trajectories were also used to
correct the TOF through the spectrometer for any given momentum
in order to  improve the time resolution. The precise path length for
each particle was also needed in order to calculate the pion-decay
correction, which is very significant at such low energies.
The resulting focal-plane spectra
were then used to extract the areas under the peaks due to the elastic
scattering with the help of Gaussian fits, including  Landau tails.
The normalisation of the  beam was provided both by the 
muon telescope ring
and by the hodoscope. Stability of the former was monitored by
comparing the ratios of counts for the four different sides to their
sum. The hodoscope provided the muon to pion ratios and it also enabled
us to normalize muon runs directly on the muons of the beam.

\begin{figure}
\epsfig{file=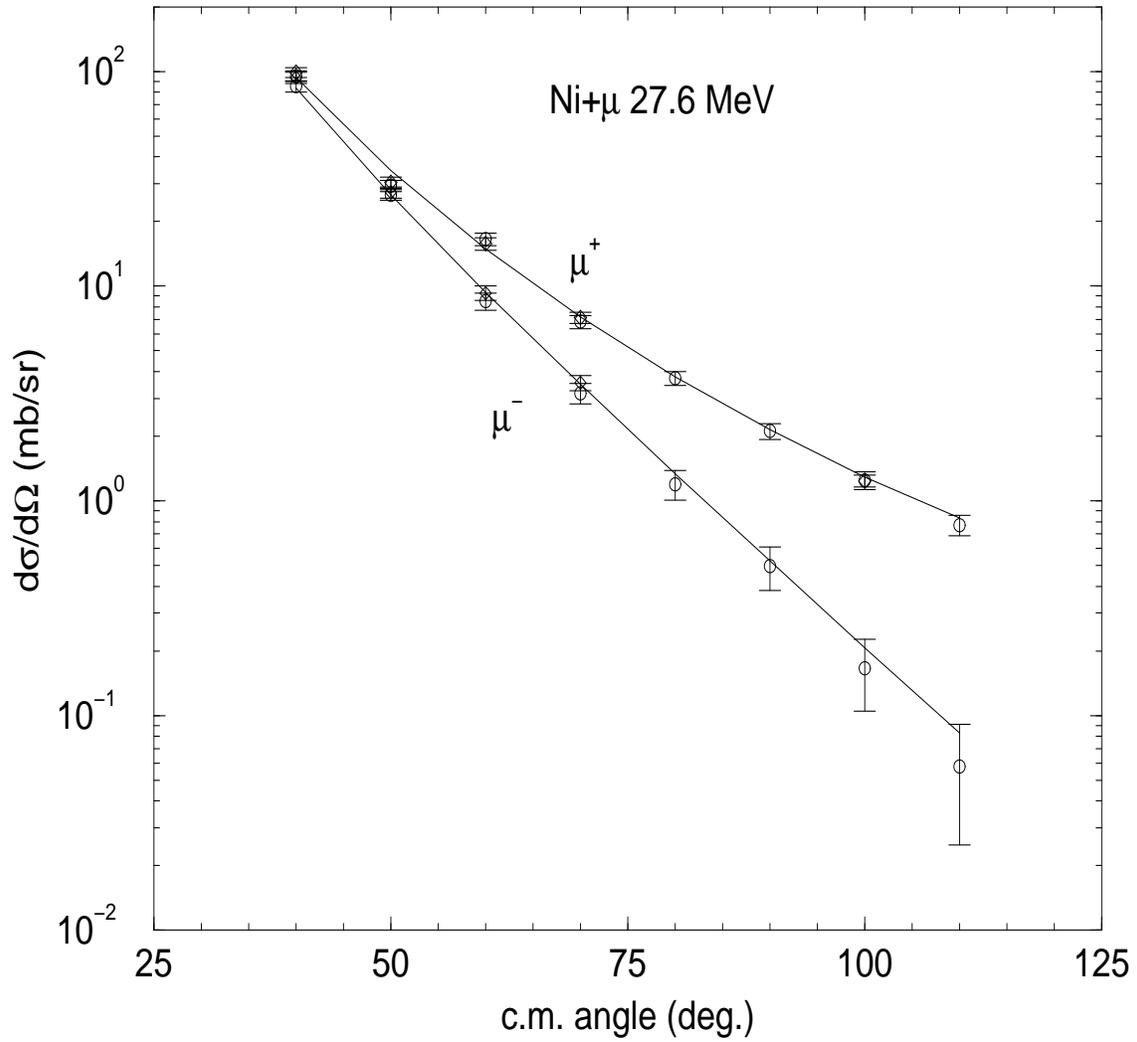, height=140mm,width=150mm}
\caption{Coulomb scattering of muons by Ni. Open circles: `self' method.
open diamonds: `muon runs' method.
Continuous curves are calculated Coulomb scattering for the finite size
charge distribution. Common normalization  constants have been
used, separately for all the $\mu^+$ points and all the $\mu^-$
points, see text.}
\label{fig:muons}
\end{figure}

The overall consistency and reliability of the determination of 
differential cross sections was demonstrated by the elastic
scattering of muons. In the `self' method 
results were available for all angles where pions were measured and
the muon cross sections were
rescaled to the pion position in the focal plane. In the `muon runs'
method the muon peak was already at the normal location of the pion
peak and no rescaling was needed. The statistics was better for
these runs but not all angles were measured. The relative cross sections
were normalized at each angle by comparing measurements to predictions 
for Coulomb scattering from the known charge distributions of the 
target nuclei \cite{FBH95} using the computer program HADES \cite{HADES}
which was modified to describe elastic muon scattering as well. 
The uncertainty in the beam momentum causes uncertainties of 1.2 to 2.0\%
in the calculated $\mu ^+$ cross sections, depending on angle, and
uncertainties of 1.3 to 2.7\% in the calculated $\mu ^-$ cross sections.
The uncertainties due to errors in the parameters of the nuclear charge
distributions are smaller than 0.5\%.
The statistical errors of the `self' method were in the range of 2 to 8\%
for $\mu ^+$ and 3.5 to 15\% for most of the $\mu ^-$ measurements. For the
`muon runs' method the corresponding errors were 2\% for $\mu ^+$
and 2 to 3\% for $\mu ^-$. The weighted average normalisation constants
for the two methods differed by 2\% when their estimated errors were
7\% and 5\%, respectively.
Consequently a common normalisation constant
was chosen as the weighted average of both methods, with an estimated
uncertainty of 4\%.
For 30$^\circ$  a separate
constant was used due to different setting of the channel slits.
Figure \ref{fig:muons} shows as an example comparisons between calculations
and measurements for the Coulomb scattering of muons by Ni.
Open circles and diamonds are for the two types of muon measurements,
`self' and `muon runs', respectively. The 30$^\circ$ points are not
included because they do not have the same normalization constant 
as discussed above.

The pion cross sections were derived using the above average muon 
normalisations. There were insignificant differences between results based on
the muon telescope ring and results based on the hodoscope for relative
normalisation of the beam. The experimental muon to pion ratios
were the averages taken from the hodoscope and corrected for the 
genuine ratios on the target, as described above. The nominal
errors included the statistical errors and the errors on the average
muon normalisation, which include also errors due to the focal plane
position dependence of the acceptance of the spectrometer. Typical
combined errors were in the range of 4-5\% and although a careful
analysis did not reveal any systematic effects, we chose to add
quadratically an estimated normalisation error of 5\% 
as done before with this spectrometer \cite{JMJ95}, when the muon normalisation
method was compared with more conventional methods.
These additional errors are
included in the final errors presented here. Note that in the previous 
$\chi ^2$ fits \cite{FBB04} it was found that the derived parameters
of the optical potential did not depend on the precise values of the 
additional normalisation errors.

The experimental results are summarized in Tables \ref{tab:Si} to
\ref{tab:Zr}. Because the target angle relative to the beam was always
half of the spectrometer angle, the mid-target energy varied  with
angle in a range of up to 0.25 MeV . The cross sections presented 
were slightly corrected for these variations
with the help of an optical model (see below) to a common energy for each
target, which corresponds to the middle of that range.

\section{Analysis}
\label{sec:analysis}

The analysis was performed with the conventional pion-nucleus optical
potential as is widely used for analysing pionic atom data \cite{BFG97}.
The Klein-Gordon equation is  written as follows:

\begin{equation} \label{equ:KGS}
\left[ \nabla^2 + k^2 - 2\varepsilon^{(A)}_{red} (V_{{\rm opt}} + V_c )+
V^2_c\right] \psi = 0~~ ~~(\hbar = c = 1)
\end{equation}

\noindent
where $k$ and $\varepsilon^{(A)}_{red}$ are the
wave number and reduced energy respectively in the c.m.
system, $(\varepsilon^{(A)}_{red})^{-1}=E_p^{-1}+E_t^{-1}$ in
terms of the c.m. energies for the projectile and target
particles, respectively.
$V_c$ is the finite-size
Coulomb interaction of the pion with the nucleus, including
vacuum-polarization terms.
The optical potential for low energy pions is the well-known potential
given by Ericson and Ericson \cite{EEr66},

\begin{equation} \label{equ:EE1}
2\varepsilon^{(A)}_{red} V_{{\rm opt}}(r) = q(r) + \vec \nabla \cdot 
\alpha(r) \vec \nabla
\end{equation}
with
\begin{eqnarray} \label{equ:EE1s}
q(r) & = & -4\pi(1+\frac{\varepsilon^{(A)}_{red}}{M})\{{\bar b_0}(r)
[\rho_n(r)+\rho_p(r)]
  +b_1[\rho_n(r)-\rho_p(r)] \} \nonumber \\
 & &  -4\pi(1+\frac{\varepsilon^{(A)}_{red}}{2M})4B_0\rho_n(r) \rho_p(r) ,
\end{eqnarray}

\begin{equation} \label{equ:LL2}
\alpha (r) = \frac{\alpha _1(r)}{1+\frac{1}{3} \xi \alpha _1(r)}
 + \alpha _2(r) ,
\end{equation}
\noindent
where
\begin{equation} \label{equ:alp1}
\alpha _1(r) = 4\pi (1+\frac{\varepsilon^{(A)}_{red}}{M})^{-1} 
\{c_0[\rho _n(r)
  +\rho _p(r)] +  c_1[\rho _n(r)-\rho _p(r)] \} ,
\end{equation}

\begin{equation} \label{equ:alp2}
\alpha _2(r) = 4\pi (1+\frac{\varepsilon^{(A)}_{red}}{2M})^{-1} 
4C_0\rho _n(r) \rho _p(r).
\end{equation}
\noindent
In these expressions $\rho_n$ and $\rho_p$ are the neutron and proton density
distributions normalized to the number of neutrons $N$ and number
of protons $Z$, respectively, 
 $M$ is the mass of the nucleon,
 $q(r)$ is referred to
as the $s$-wave potential term and $\alpha(r)$ is referred to
as the $p$-wave potential term.
The function ${\bar b_0}(r)$ in Eq. (\ref{equ:EE1s})
is given in terms of the {\it local} Fermi
momentum $k_{\rm F}(r)$ corresponding to the isoscalar nucleon
density distribution:
\begin{equation} \label{equ:b0b}
{\bar b_0}(r) = b_0 - \frac{3}{2\pi}(b_0^2+2b_1^2)k_{\rm F}(r),
\end{equation}
where $b_0$ and $b_1$ are minus the pion-nucleon isoscalar
and isovector effective scattering lengths, respectively.
The quadratic terms in $b_0$ and $b_1$ represent double-scattering
modifications of $b_0$. In particular, the $b_1^2$ term represents
a sizable correction to the nearly vanishing linear $b_0$ term.
The  coefficients $c_0$ and
$c_1$ in Eq. (\ref{equ:alp1}) are the pion-nucleon
isoscalar and isovector $p$-wave scattering volumes,
respectively. The parameters $B_0$ and $C_0$
in Eqs. (\ref{equ:EE1s}) and (\ref{equ:alp2})
represent $s$-wave and $p$-wave
absorption, respectively, on pairs
of nucleons and as such they have imaginary parts.
Dispersive real parts are found
to play an important role in pionic atom potentials.
The terms with $4\rho _n \rho _p$ were originally written as
$(\rho _n+\rho _p)^2$, but the results hardly
depend on which form is used.
The parameter $\xi$ in Eq. (\ref{equ:LL2}) is the usual
Ericson-Ericson Lorentz-Lorenz (EELL) coefficient  \cite{EEr66}.
An additional relatively small term,
known as the `angle-transformation' term (see Eq.(24) of \cite{BFG97}),
is also included.

For pionic atoms the scattering lengths ($b_0$ and $b_1$) and scattering
volumes ($c_0$ and $c_1$) are real but for the scattering case they 
become complex. At 21 MeV the imaginary parts of these parameters for
the free pion-nucleon interaction are small, and in the present analysis,
in order not to increase further the number of parameters,
we assume  only $B_0$ and $C_0$ to be complex. Note that in the preliminary
analysis of the present data \cite{FBB04} we used complex $c_0$ and $c_1$
because we observed the need to make the $p$-wave absorption 
charge-dependent. Here we adhere to the pionic atoms approach but
allow also such charge-dependence. The proton density distribution
$\rho _p$ was obtained from the known charge density of the target nucleus
by unfolding the finite size of the proton. For the neutrons we used
an `average' shape \cite{FGa04a} with rms radii differences of 
$r_n-r_p$=$-$0.04, $-$0.05, 0.01 and 0.12 fm for Si, Ca, Ni and Zr,
respectively. Calculations were made using for the densities 
the natural isotopic mixture of the
targets as it was found to be an excellent approximation to calculations
done separately for the various isotopes and then averaged.

Fits to the data were made by the usual least-squares method, handling
all 72 cross sections at the same time. With at least 8 parameters in
the potential and with
correlations between some parameters it was always possible to get 
excellent fits with $\chi ^2$
per degree of freedom very close to 1.0. In such circumstances it is necessary
to apply constraints and physical considerations in order to get
meaningful results. The rest of this section is devoted to results
obtained by adopting this approach. Note, however, that in global
analyses of pionic atom data covering the whole of the periodic table,
it is found ({\it e.g.} in Ref. \cite{Fri02a}) that all the parameters
are quite well-determined.

First it was noticed that the $p$-wave scattering volumes
$c_0$ and $c_1$ always came out very close to their corresponding free
pion-nucleon values of 0.21 and 0.165 $m_\pi ^{-3}$, respectively.
Consequently, $c_0$ and $c_1$ were kept fixed at these values in 
subsequent fits.
The EELL coefficient $\xi$ was also kept constant at 1.0.
Without any additional restrictions on parameter values, it was found that
the best fit was achieved with a very repulsive real part for the
$s$-wave potential. That was clearly seen with $b_0$ being  a factor 2-3
too repulsive compared to the free $\pi$N amplitude and with the dispersive
Re$B_0$ approximately equal to $-$4Im$B_0$. Both results represent 
`anomalously'  large repulsion,
which is parallel to the `anomalous repulsion' in pionic atoms.

Next we turn to the isovector $s$-wave parameter $b_1$ which has been the
topic of interest in recent years, as outlined in the Introduction. 
With the conventional model for the potential (denoted hereafter by C) $b_1$ 
is assumed to be constant, and then we get (see table \ref{tab:potls})
 $b_1=-0.114\pm0.006~m_\pi ^{-1}$, compared to the free pion-nucleon value
\cite{SAID}  of $-0.088\pm0.001~ m_\pi ^{-1}$, 
clearly exhibiting an `anomalous'
repulsion. In the Weise model \cite{Wei01}
the in-medium $b_{1}$ is related to
possible partial restoration of chiral symmetry in dense matter,
as follows.
Since $b_{1}$ in free-space is well approximated in lowest
chiral-expansion order by the Tomozawa-Weinberg expression
\cite{Tom66}
\begin{equation}
\label{equ:b1}
b_{1}=-\frac{\mu_{\pi N}}{8 \pi f^{2}_{\pi}}=-0.08~m^{-1}_{\pi} \,,
\end{equation}
then it can be argued that $b_{1}$ will be modified in pionic atoms
if the pion decay constant
$f_\pi$ is modified in the medium. The square of this decay constant
is given, in leading order,
 as a linear function of the nuclear density,
\begin{equation} \label{eq:fpi2}
f_\pi ^{*2} = f_\pi ^2 - \frac{\sigma }{m_\pi ^2} \rho
\end{equation}
with $\sigma$ the pion-nucleon sigma term.
This leads to a density-dependent isovector amplitude such that $b_1$ becomes

\begin{equation}\label{eq:ddb1}
b_1(\rho) = \frac{b_1(0)}{1-2.3\rho}
\end{equation}
for $\sigma $=50 MeV \cite{GLS91} and
with $\rho$ in units of fm$^{-3}$. This model which was shown
\cite{Fri02,Fri02a}
to explain the anomaly in pionic atoms, is denoted here by W.
An alternative approach which was also shown to explain the anomaly
in pionic atoms \cite{FGa04} is to replace $E$ by $E-V_c$ 
by imposing the minimal substitution requirement \cite{ETa82}, which
is effective through the energy dependence of the isoscalar parameter $b_0$.
This model is denoted here by E. Finally, applying both mechanisms we
have the EW model.

\begin{figure}
\epsfig{file=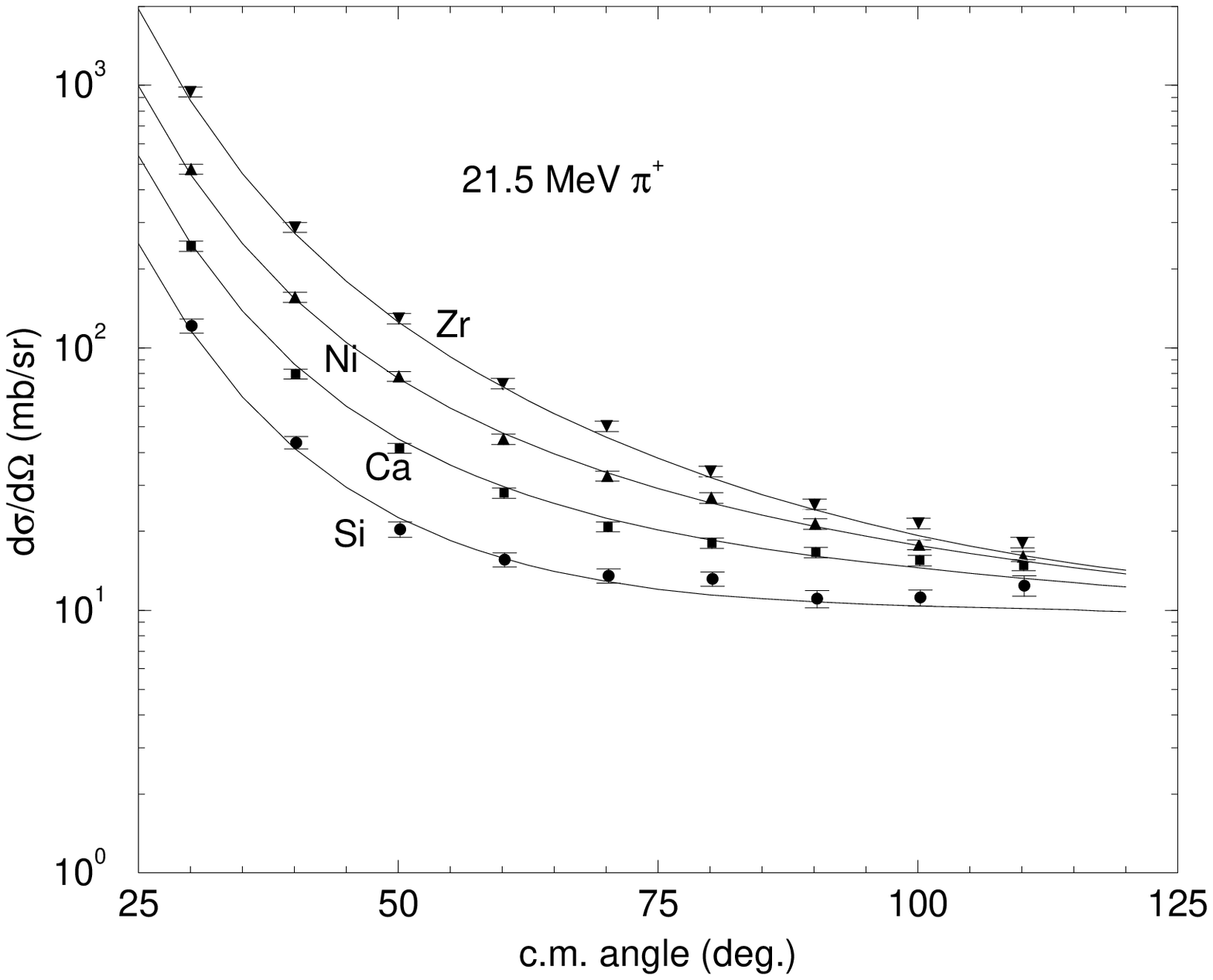, height=140mm,width=150mm}
\caption{Experimental results for $\pi ^+$ compared to predictions 
for potential EW of Table \ref{tab:potls}. Potential parameters
were obtained from fits to the combined data for $\pi ^+$ and $\pi ^-$.}
\label{fig:pions+}
\end{figure}

\begin{figure}
\epsfig{file=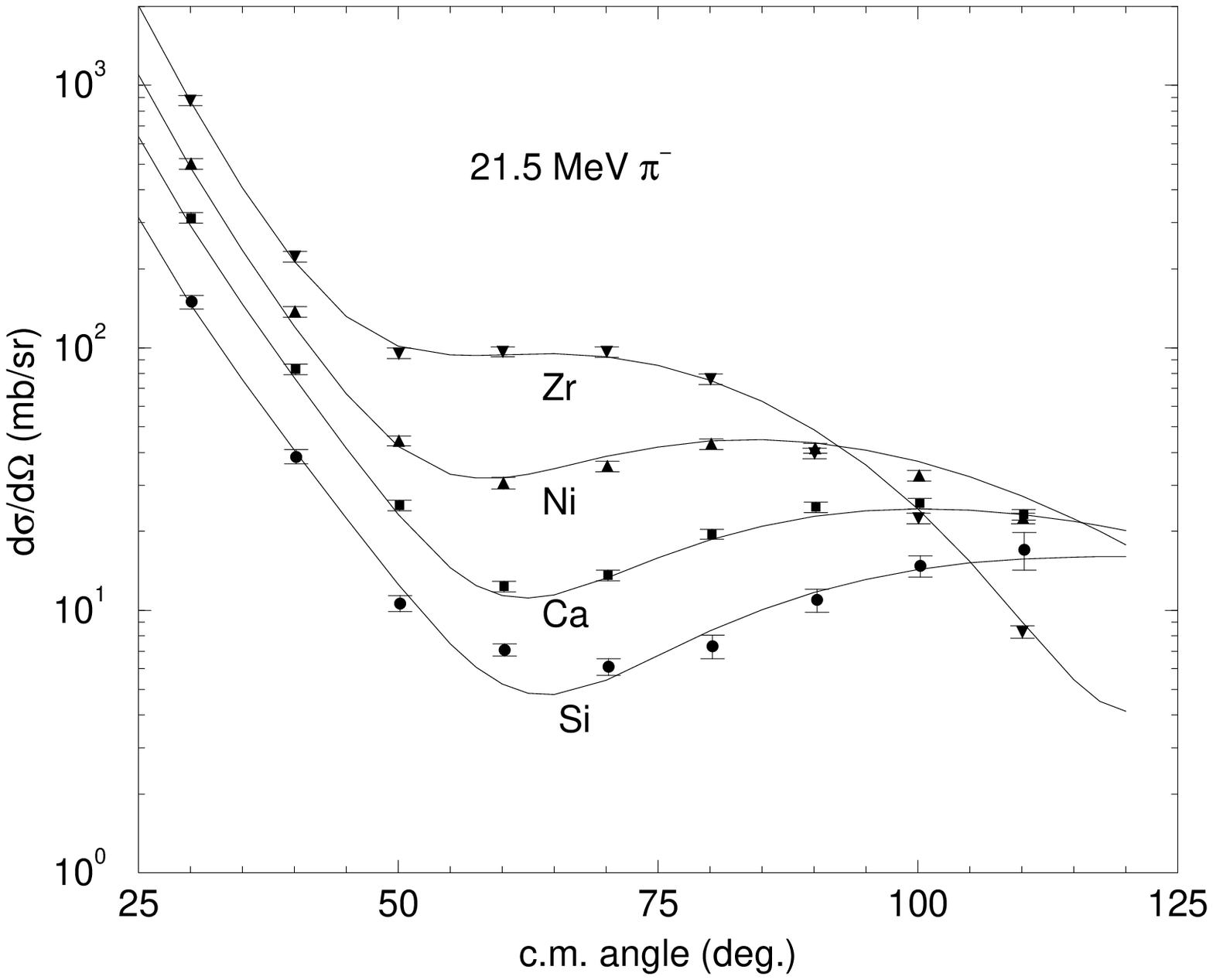, height=140mm,width=150mm}
\caption{Experimental results for $\pi ^-$ compared to predictions
for potential EW of Table \ref{tab:potls}. Potential parameters
were obtained from fits to the combined data for $\pi ^+$ and $\pi ^-$.}
\label{fig:pions-}
\end{figure}

Table \ref{tab:potls} summarizes the parameters of the optical potentials
obtained for the various models with the constraints of (i) $b_0$ in the 
range \cite{SAID} of $-0.008$ to $-0.010~m_\pi ^{-1}$ and 
(ii) Re$B_0$ close to $-$Im$B_0$ (see Ref.\cite{BFG97}). For 
the empirical parameter
Im$C_0$ we introduced the option of energy dependence, implemented as

\begin{equation}\label{eq:Coul}
{\rm Im}C_0~=~{\rm Im}C_0 ^0~+\alpha V_c
\end{equation}
but the parameter $\alpha$ turned out to be consistent with zero except
for the EW model. The last row in the table, marked as EW$^*$, is for
the potential EW with this energy dependence of Im$C_0$. The slope parameter
in that case is $\alpha=-0.003\pm0.001~m_\pi ^{-6}$ MeV$^{-1}$, which implies 
good consistency between pionic atoms and the potential at 21 MeV.
It is seen from Table \ref{tab:potls} that the fit due to the conventional
potential (C) is very significantly improved by introducing the density
dependence of $b_1$, the energy dependence of $b_0$ or both. The derived
values of $b_1$ agree with the free pion-nucleon value of 
$-0.088\pm0.001~ m_\pi ^{-1}$ only when the density dependence is 
included in the model. Figures \ref{fig:pions+} and \ref{fig:pions-}
show comparisons between experiment and calculation for the EW potential.
It is self evident that the calculations reproduce all the features of 
the data.

The success of the W model is easy to understand since the density
dependence introduced by Weise \cite{Wei01} provides enhanced $s$-wave
repulsion in the nuclear medium. The observation that the
agreement between calculation
and experiment improves when the energy dependence of the parameter $b_0$
is included is also easy to understand. Out of the scattering lengths and 
scattering volumes for the free $\pi N$ interaction at low energies the
isoscalar scattering length $b_0$ has the strongest energy dependence
\cite{SAID}, and coupled with its very small values that energy dependence
is significant. By adopting the minimal substitution requirement \cite{ETa82} 
of $E \to E - V_{c}$ and using the empirical slope of $b_0(E)$, the effects
of the Coulomb attraction or repulsion are taken into account. In the
scattering scenario it causes the 
{\it effective} values of $b_0$ to be different for
the two charge states of the pion, and that is favoured by the data. 
The quadratic terms of the potential, namely, $B_0$ and $C_0$, remain
empirical and are determined to a rather low accuracy. Only for the EW*
model there is an indication that Im$C_0$ might be energy dependent.

\section{Summary}
\label{sec:sum}
Precision measurements of elastic scattering of 21.5 MeV $\pi ^\pm$ 
by targets of Si, Ca, Ni and Zr have been made in order to provide
an experimental basis for determining
 parameters of the pion-nucleus
optical potential. Emphasis was placed on the `anomalous' $s$-wave
repulsion, well known from global fits to pionic atom data.
 The experiment used a single arm magnetic
spectrometer which enabled efficient particle identification and
rejection of the large muon background, typical of such low energies.
The absolute scale and the validity of the shapes of angular distributions
measured for pions were established by parallel measurements of the
Coulomb scattering of muons by the same target nuclei. 
The data were analyzed with optical potentials known from fits
to pionic atom data.
Very good fits to the elstic scattering data
were obtained with extension to 21 MeV of several of the commonly 
accepted pionic atoms potentials. The `anomalous' $s$-wave
repulsion was clearly observed, and could be accounted for by introducing
a chiral-motivated density dependence of the isovector scattering amplitude,
as is the case with pionic atoms. This density dependence that
enhances the isovector $s$-wave amplitude in the nuclear medium, 
also improves significantly the fit to the data. 
Introducing into the potential the empirical energy dependence 
of the $s$-wave  isoscalar 
amplitude, motivated by the minimal substitution requirement 
of $E \to E - V_{c}$ and using the empirical slope of $b_0(E)$,
 greatly improves the fits to the data. However, this effect
on its own is incapable of explaining the `anomaly' at 21 MeV, {\it unlike} 
the situation with pionic atoms. 
In all cases the linear terms of the pion-nucleus potential connect
smoothly to the corresponding quantities obtained from fits to pionic
atom data. In contrast the empirical quadratic terms are quite different
from the corresponding values for pionic atoms. 
However, when both the density 
dependence and the energy dependence 
mechanisms are included in the model, it is found that
making the $p$-wave absorption term energy-dependent, via the Coulomb
potential, further improves the fits to the data. In that case the
$p$-wave absorption term also connects smoothly to the pionic atoms value.

\vspace{1cm}
This work was supported in part by
the German ministry of education and research (BMBF) under contracts
06 TU 987I and  06 TU 201
and the Deutsche Forschungsgemeinschaft (DFG) through European Graduate
School 683 and Heisenberg Program.

\begin{table}
\caption{Experimental results for 21.69 MeV $\pi ^{\pm}$ scattering from Si.}
\label{tab:Si}
\begin{tabular}{ccc}
c.m. angle ($^\circ$)&$\pi^+ ~~d\sigma/d\Omega$ (mb/sr) &
$\pi^- ~~d\sigma/d\Omega$ (mb/sr)
\\ \hline 
30.15 & 121.4 $\pm$ 9.6 & 149.7 $\pm$   11.7\\
40.19 & 43.5  $\pm$   3.2  & 38.5 $\pm$    3.1 \\
 50.23 &  20.4 $\pm$    1.7 & 10.6 $\pm$    0.9 \\
 60.25 &   15.6  $\pm$   1.2 & 7.1  $\pm$   0.5 \\
 70.28  &  13.5 $\pm$    1.1 &   6.1 $\pm$    0.5 \\  
 80.29  &  13.2 $\pm$    1.0 & 7.3  $\pm$   0.8 \\
 90.30  &  11.1 $\pm$    1.0 & 10.9 $\pm$    1.2  \\
 100.29 &   11.2 $\pm$    1.0 &  14.7 $\pm$    1.6 \\
 110.28 &   12.4 $\pm$    1.3 &  17.0 $\pm$    2.9 \\
\end{tabular}
\end{table}

\begin{table}
\caption{Experimental results for 21.05 MeV $\pi ^{\pm}$ scattering from Ca.}
\label{tab:Ca}
\begin{tabular}{ccc}
c.m. angle ($^\circ$)&$\pi^+ ~~d\sigma/d\Omega$ (mb/sr) &
$\pi^- ~~d\sigma/d\Omega$ (mb/sr)
\\ \hline
30.11 &  243.9 $\pm$   16.6 &  312.7  $\pm$  21.1 \\
40.13 &   79.6 $\pm$    5.4 &  82.9   $\pm$  5.7  \\
50.16 &   41.4 $\pm$    2.8 &  25.1   $\pm$  1.7  \\
60.18 &   28.1 $\pm$    1.9 &  12.3   $\pm$  0.8  \\
70.19 &   20.8 $\pm$    1.4 &  13.6   $\pm$  0.9  \\
80.20 &   18.0 $\pm$    1.2 &  19.5   $\pm$  1.3  \\
90.21 &   16.6 $\pm$    1.1 &  24.7   $\pm$  1.7  \\
100.20 &     15.5 $\pm$     1.1 & 25.6   $\pm$  1.7  \\
110.20 &    14.9 $\pm$    1.0  & 23.2   $\pm$  1.6  \\
\end{tabular}
\end{table}

\begin{table}
\caption{Experimental results for 21.45 MeV $\pi ^{\pm}$ scattering from Ni.}
\label{tab:Ni}
\begin{tabular}{ccc}
c.m. angle ($^\circ$)&$\pi^+ ~~d\sigma/d\Omega$ (mb/sr) &
$\pi^- ~~d\sigma/d\Omega$ (mb/sr)
\\ \hline
30.07 &  479.5 $\pm$   31.2 & 502.6  $\pm$  33.7  \\
40.09 &  156.0 $\pm$   10.4 & 137.1  $\pm$   9.2 \\
50.11 &   77.8 $\pm$    5.2 & 44.2   $\pm$  3.0 \\
60.12 &   44.9 $\pm$    3.0 & 30.5   $\pm$  2.1 \\
70.13 &   32.5 $\pm$    2.2 & 35.4   $\pm$  2.4 \\
80.14 &   26.8 $\pm$    1.8 & 43.0   $\pm$  2.9 \\
90.14 &   21.3 $\pm$    1.4 & 41.5   $\pm$  2.8 \\
100.14 &    17.8 $\pm$    1.2 &  32.6  $\pm$   2.2\\ 
110.13 &    16.0 $\pm$    1.1 & 22.4   $\pm$  1.5 \\
\end{tabular}
\end{table}

\begin{table}
\caption{Experimental results for 21.70 MeV $\pi ^{\pm}$ scattering from Zr.}
\label{tab:Zr}
\begin{tabular}{ccc}
c.m. angle ($^\circ$)&$\pi^+ ~~d\sigma/d\Omega$ (mb/sr) &
$\pi^- ~~d\sigma/d\Omega$ (mb/sr)
\\ \hline
30.05 &  943.6  $\pm$  62.1  & 875.4 $\pm$   58.8  \\
40.06 &  287.9  $\pm$  19.5 & 222.5  $\pm$  15.1 \\
50.07 &  129.3  $\pm$   8.7 & 95.2   $\pm$  6.5 \\
60.08 &   73.0  $\pm$   4.9 & 96.7   $\pm$  6.5 \\
70.09 &   50.3  $\pm$   3.4 & 96.5   $\pm$  6.5 \\
80.09 &   33.9  $\pm$   2.3 & 75.9   $\pm$  5.1 \\
90.09 &   25.3  $\pm$   1.7 & 39.7   $\pm$  2.7 \\
100.09 &   21.5 $\pm$    1.5 &  22.4 $\pm$    1.5  \\
110.09 &   18.1 $\pm$    1.2 &  8.3  $\pm$   0.6 \\
\end{tabular}
\end{table}

\begin{table}
\caption{Optical potential parameters from fits to elastic scattering
of 21.5 MeV $\pi ^\pm$ by Si, Ca, Ni and Zr. Values of $b_0$ were kept
in the range of $-$0.008  to  $-$0.010 $m_\pi ^{-1}$ and values of
$c_0$ and $c_1$ were held fixed at the free pion-nucleon values
of 0.21 and 0.165 $m_\pi ^{-3}$, respectively. Values of Re$B_0$ were
held fixed. For the EW$^*$ model Coulomb dependence of Im$C_0$ was
included, see Eq. (\ref{eq:Coul}). The various models are defined
in the text} 
\label{tab:potls}
\begin{tabular}{ccccccc}
model &$b_1 (m_\pi ^{-1}$)&Re$B_0 (m_\pi ^{-4}$)&Im$B_0 (m_\pi ^{-4}$)&
Re$C_0( m_\pi ^{-6}$)&Im$C_0( m_\pi ^{-6}$)& $\chi ^2$ for 72 points \\
\\ \hline
C &$-$0.114$\pm$0.006&\underline{$-$0.040}&0.035$\pm$0.007&$-$0.030$\pm$0.020&
0.075$\pm$0.035&134 \\
W &$-$0.081$\pm$0.005&\underline{$-$0.040}&0.040$\pm$0.007&$-$0.045$\pm$0.015&
0.062$\pm$0.030&88 \\
E &$-$0.119$\pm$0.006&\underline{$-$0.025}&0.025$\pm$0.005&$-$0.022$\pm$0.012&
0.085$\pm$0.027&80 \\
EW&$-$0.083$\pm$0.005&\underline{$-$0.025}&0.023$\pm$0.005&$-$0.023$\pm$0.012&
0.130$\pm$0.025&88 \\
EW$^*$&$-$0.083$\pm$0.005&\underline{$-$0.025}&0.018$\pm$0.004&$-$0.010$\pm$0.009&
0.140$\pm$0.020&72 \\
\end{tabular}
\end{table}

\begin{thebibliography}{xyz03}





\bibitem{BFG97}For early references see  C.J. Batty, E. Friedman and A. Gal,
Phys. Rep. {\bf 287}, 385 (1997).


\bibitem{YHI96}T. Yamazaki, R.S. Hayano, K. Itahashi, K. Oyama, A. Gillitzer,
H. Gilg, M. Kn\"{u}lle, M. M\"{u}nch, P. Kienle, W. Schott, H. Geissel,
N. Iwasa and G. M\"{u}nzenberg, Z. Phys. A {\bf 355}, 219 (1996).

\bibitem{GGK00}H. Gilg, A. Gillitzer, M. Kn\"ulle, M. M\"unch, W. Schott,
P. Kienle, K. Itahashi, K. Oyama, R.S. Hayano, H. Geissel, N. Iwasa,
G. M\"unzenberg and T. Yamazaki,  Phys. Rev. C {\bf 62}, 025201 (2000).

\bibitem{GGG02}H. Geissel, H. Gilg, A. Gillitzer, R.S. Hayano, S. Hirenzaki,
K. Itahashi, M. Iwasaki, P. Kienle, M. M\"unch, G. M\"unzenberg, W. Schott,
K. Suzuki, D. Tomono, H. Weick, T. Yamazaki and T. Yoneyama, Phys. Rev.
Lett. {\bf 88}, 122301 (2002).

\bibitem{SFG04}K. Suzuki, M. Fujita, H. Geissel, H. Gilg, A. Gillitzer,
R.S. Hayano, S. Hirenzaki, K. Itahashi, M. Iwasaki, P. Kienle, M. Matos,
G. M\"unzenberg, T. Ohtsubo, M. Sato, M. Shindo, T. Suzuki, H. Weick,
M. Winkler, T. Yamazaki and T. Yoneyama, Phys Rev. Lett. {\bf 92}, 
072302 (2004).


\bibitem{FSo85}E. Friedman and G. Soff,
J. Phys. G: Nucl. Phys. {\bf 11}, L37 (1985).

\bibitem{TYa88}H. Toki and T. Yamazaki, Phys. Lett. B {\bf 213}, 129 (1988).

\bibitem{THY89}H. Toki, S. Hirenzaki, T. Yamazaki and R.S. Hayano,
Nucl. Phys. A {\bf 501}, 653 (1989).

\bibitem{Wei01}W. Weise, Nucl. Phys. A {\bf 690}, 98c (2001).

\bibitem{KWe01}N. Kaiser and W. Weise, Phys. Lett. B {\bf 512}, 283 (2001).

\bibitem{MOW02}U.G. Meissner, J.A. Oller and A. Wirzba, Ann. Phys. (NY)
{\bf 297}, 27 (2002).

\bibitem{KKW03}E.E. Kolomeitsev, N. Kaiser and W. Weise,  Phys. Rev.
Lett. {\bf 90}, 092501 (2003).

\bibitem{KKW03a}E.E. Kolomeitsev, N. Kaiser and W. Weise,
 Nucl. Phys. A {\bf 721}, 835c (2003).


\bibitem{Fri02}E. Friedman, Phys. Lett. B {\bf 524}, 87 (2002).

\bibitem{Fri02a}E. Friedman, Nucl. Phys. A {\bf 710}, 117 (2002).

\bibitem{KYa01}P. Kienle and T. Yamazaki, Phys. Lett. B {\bf 514}, 1 (2001).

\bibitem{GGG02a}H. Geissel et al.,
H. Gilg, A. Gillitzer, R.S. Hayano, S. Hirenzaki,
K. Itahashi, M. Iwasaki, P. Kienle, M. M\"unch, G, M\"unzenberg, W. Schott,
K. Suzuki, D. Tomono, H. Weick, T. Yamazaki and T. Yoneyama,
Phys. Lett. B {\bf 549}, 64 (2002).

\bibitem{FGa03}E. Friedman and A. Gal, Nucl. Phys. A {\bf 724}, 143 (2003).

\bibitem{FGa98}E. Friedman and A. Gal, Phys. Lett. B {\bf 432}, 235 (1998).

\bibitem{FGa03a}E. Friedman and A. Gal, Nucl. Phys. A {\bf 721}, 842c (2003).

\bibitem{FGa04}E. Friedman and A. Gal, Phys. Lett. B {\bf 578}, 85 (2004).

\bibitem{ETa82}T.E.O. Ericson and L. Tauscher, Phys. Lett. B {\bf 112}, 
425 (1982).

\bibitem{Wri88}D.H. Wright,M. Blecher, B. G. Ritchie, D. Rothenberger,
R. L. Burman, Z. Weinfeld, J. A. Escalante, C. S. Mishra, 
and C. S. Whisnant,    
Phys. Rev. C {\bf 37},  1155 (1988).


\bibitem{FBB04}E. Friedman, M. Bauer, J. Breitschopf,
H. Clement, H. Denz, E. Doroshkevich,
A. Erhardt,  G.J. Hofman,
R. Meier, G.J.Wagner, G. Yaari, Phys. Rev. Lett. {\bf 93}, 122302 (2004).

\bibitem{For97}F. Foroughi (1997) http://people.web.psi.ch/foroughi/.

\bibitem{MGJ87}H.Matth\"ay, K.G\"oring, J.Jaki, W.Kluge, M.Metzler and
U. Wiedner, Proc. Int. Symposium on {\it Dynamics of Collective
Phenomena in Nuclear and Subnuclear Long Range Interactions in
Nuclei},
Bad Honnef, Germany, May 4-7 1987, ed. Peter David, World Scientific,
Singapore, New Jersey, Hong Kong.

\bibitem{JMJ95}Ch. Joram, M. Metzler, J. Jaki, W. Kluge, H. Matth\"ay,
R. Wieser, B.M. Barnett, H. Clement, S. Krell and G.J. Wagner,
Phys. Rev. C {\bf 51}, 2144 (1995).

\bibitem{BKC90}B.M. Barnett, S. Krell, H. Clement, G.J. Wagner, J. Jaki,
Ch. Joram, W. Kluge, H. Matth\"ay and M. Metzler, 
Nucl. Inst. Methods  A {\bf297}, 444 (1990).

\bibitem{FBH95}G. Fricke, C. Bernhardt, K. Heilig, L.A. Schaller,
L. Schellenberg, E.B. Shera, C.W. De Jager, At. Data Nucl. Data
Tables {\bf 60}, 177 (1995).

\bibitem{HADES} H.G.Andresen, H. Peter, M. M\"uller, H.J.Ohlbach and P.Weber,
Program HADES, Universit\"at Mainz (1986), unpublished.

\bibitem{EEr66} M. Ericson, T.E.O. Ericson, Ann. Phys. (NY) {\bf 36},
323 (1966).

\bibitem{FGa04a} E. Friedman, A. Gal, Nucl. Inst. Methods B {\bf 214},
160 (2004).


\bibitem{SAID}http://gwdac.phys.gwu.edu/


\bibitem{Tom66}Y. Tomozawa, Nuovo Cimento A {\bf 46}, 707 (1966);
S. Weinberg, Phys. Rev. Lett. {\bf 17}, 616 (1966).

\bibitem{GLS91}J.Gasser, H. Leutwyler and M.E. Sainio, Phys. Lett. B
{\bf 253}, 252 (1991).

\end{thebibliography}
\end{document}